# Heterogeneous accretion of Earth inferred from Mo-Ru isotope systematics

Timo Hopp*,1, Gerrit Budde2, and Thorsten Kleine

Institut für Planetologie, University of Münster, Wilhelm-Klemm-Str. 10, 48149 Münster, Germany.

*Corresponding author: hopp@uchicago.edu

1 Present address: Origins Laboratory, Department of Geophysical Sciences and Enrico Fermi Institute, The University of Chicago, 5743 South Ellis Avenue, Chicago, IL 60637, USA.

2 Present address: The Isotoparium, Division of Geological and Planetary Sciences, California Institute of Technology, Pasadena, CA 91125, USA.

**Abstract**

The Mo and Ru isotopic compositions of meteorites and the bulk silicate Earth (BSE) hold important clues about the provenance of Earth's building material. Prior studies have argued that non-carbonaceous (NC) and carbonaceous (CC) meteorite groups together define a Mo-Ru 'cosmic' correlation, and that the BSE plots on the extension of this correlation. These observations were taken as evidence that the final 10–15% of Earth's accreted material derived from a homogeneous inner disk reservoir with an enstatite chondrite-like isotopic composition. Here, using new Mo and Ru isotopic data for previously uninvestigated meteorite groups, we show that the Mo-Ru correlation only exists for NC meteorites, and that both the BSE and CC meteorites fall off this Mo-Ru correlation. These observations indicate that the final stages of Earth's accretion were heterogeneous and consisted of a mixture of NC and CC materials. The Mo-Ru isotope systematics are best accounted for by either an NC heritage of the late veneer combined with a CC heritage of the Moon-forming giant impactor, or by mixed NC-CC compositions for both components. The involvement of CC bodies in the late-stage accretionary assemblage of Earth is consistent with chemical models for core-mantle differentiation, which argue for the addition of more oxidized and volatile-rich material toward the end of Earth's formation. As such, this study resolves the inconsistencies between homogeneous accretion models based on prior interpretations of the Mo-Ru systematics of meteorites and the chemical evidence for heterogeneous accretion of Earth.

# 1. Introduction

The formation of Earth involved the accretion of numerous planetesimals and Moon- to Mars-sized planetary embryos. Dynamical models of planetary accretion predict that Earth's building blocks derived from a wide area of the protoplanetary disk and included some carbonaceous chondrite-like material from the outer solar system (e.g., Morbidelli et al., 2012; O'Brien et al., 2014). Consistent with this, chemical models of core-mantle differentiation show that Earth most likely formed from a heterogeneous assemblage of bodies, which were initially reduced and volatile-poor and became increasingly oxidized and volatile-rich toward the later stages of accretion (e.g., Wade and Wood, 2005; Rubie et al., 2015; Grewal et al., 2019). Until now, however, these heterogeneous accretion models have been difficult to reconcile with the isotopic record of Earth's accretion. For all elements investigated to date, Earth is isotopically most similar to enstatite chondrites, a small group of highly reduced meteorites from the inner solar system (see summary in Dauphas, 2017). This close isotopic link has led to the idea that Earth predominantly accreted from enstatite chondrite-like material (Javoy et al., 2010) or, more specifically, from a homogeneous inner disk region with an enstatite chondrite-like isotopic composition (Dauphas et al., 2014). Thus, there is an obvious discrepancy between chemical models, which argue for heterogeneous accretion, and the isotopic observations, which seem to favor homogeneous accretion of Earth.

Central to this debate are the isotope anomalies in Mo and Ru, which are important for several reasons. First, as a moderately siderophile element, the budget of Mo in the bulk silicate Earth (BSE) has been established during the last 10–15% of accretion (Dauphas, 2017). By contrast, the Ru in the BSE predominantly derives from the late veneer, i.e., the ~0.5% of broadly chondritic material added to the mantle after the cessation of core formation (e.g., Walker et al., 2015). Thus, Mo and Ru isotopes provide information on the genetic characteristics of accreted material during distinct phases of Earth's late growth history, which in turn provides a potential way of discriminating between different models of Earth's accretion (Fischer et al., 2018). Second, Mo and Ru isotope anomalies in meteorites seem to be broadly correlated, most likely reflecting the heterogeneous distribution of a common carrier of Mo and Ru produced in the *s*-process of stellar nucleosynthesis. The BSE plots at one end of this Mo-Ru 'cosmic' correlation (Dauphas et al., 2004), close to the isotopic composition of enstatite chondrites. This observation has been interpreted to indicate that the genetic characteristics of Earth's accretionary assemblage did not change during the final ~10–15% of Earth's growth and were similar to enstatite chondrites or IAB iron meteorites (Bermingham et al., 2018). Finally, for Mo (and for other elements such as Cr, Ti, and Ni) there is a fundamental isotopic dichotomy between non-carbonaceous (NC) and carbonaceous (CC) meteorites (Warren, 2011; Budde et al., 2016; Poole et al., 2017; Worsham et al., 2017, Nanne et al., 2019). These groups represent two spatially and compositionally distinct reservoirs in the early protoplanetary disk, which presumably were separated by the early formation of Jupiter (Warren, 2011; Budde et al., 2016; Kruijer et al., 2017). As such, the NC reservoir most likely represents the inner, while the CC reservoir represents the outer solar system (Warren, 2011).

Recently, Budde et al. (2019) demonstrated that the Mo isotopic composition of the BSE is intermediate between those of the NC and CC reservoirs, indicating that Earth accreted carbonaceous chondrite-like material during the late stages of its growth. Moreover, Worsham et al. (2019) have shown that CC meteorites do not plot on the Mo-Ru cosmic correlation defined by NC meteorites, implying that if Earth accreted CC material during the late stages of its growth, then the BSE should also plot off the Mo-Ru correlation. As noted above, however, current data suggest that the BSE plots on the Mo-Ru cosmic correlation (Bermingham et al., 2018).

To address these issues and better understand the Mo-Ru isotopic record of Earth's late growth history, we obtained high-precision Ru and Mo isotopic data for several previously uninvestigated meteorite groups. The new data are used to more precisely define the Mo-Ru correlation and the position of the BSE relative to this correlation, which in turn provides new constraints on the genetic characteristics of the late stage accretionary assemblage of Earth.

## 2. Samples and analytical methods

*2.1. Samples and sample preparation*

Samples selected for this study include acapulcoite-lodranites, aubrites, brachinites, mesosiderites, ureilites, winonaites, several ungrouped achondrites, and one primitive enstatite achondrite (NWA 2526). Three of the samples are classified as ungrouped achondrites (NWA 6112, NWA 5363, NWA 1058), but their chemical and petrologic properties suggest they may be linked to brachinites. Accordingly, these meteorites are sometimes also classified as 'brachinite-like' meteorites (Day et al., 2012; Hasegawa et al., 2019).

Pieces of meteorites (~1-3 g) were cut from larger slices using a diamond saw, carefully cleaned by polishing with SiC abrasives as well as sonication in ethanol, and then manually ground to a fine bulk powder in an agate mortar. For the Norton County aubrite, ~0.5 g (Ru isotope analysis) and ~0.9 g (Mo isotope analysis) pieces of a larger metal nodule (~7 g) were used. For the Ru isotope analyses, ~0.2–1 g of each sample powder was digested using inverse *aqua regia* inside sealed Carius tubes at 220 °C for 2 days (Shirey and Walker, 1995). This method is not capable of fully digesting presolar grains (Fischer-Gödde et al., 2015), but the samples of this study were all subjected to high temperatures on their parent bodies and so no longer contain presolar grains. After digestion, the sample solutions were transferred into 50 ml centrifuge tubes and centrifuged for 20 minutes to separate and remove the un-dissolved silicates. The supernatant was then dried down at 100 °C in 15 ml Savillex PFA vials, followed by repeated dry downs with 6 M HCl.

For the Mo isotope analyses, ~0.5–1 g of the sample powders was digested in Savillex vials on a hot plate using $HF-HNO_3-HClO_4$ (2:1:0.01) at 180–200 °C, followed by inverse *aqua regia* at 130-150 °C and repeated dry downs with 6 M HCl-0.06 M HF (Budde et al., 2018).

In addition to the bulk meteorites, the Ru isotopic composition of an acid leachate of the unequilibrated ordinary chondrite NWA 2458 (L3.2) was analyzed. This leachate was originally prepared for a previous study, which also reports Mo isotopic data for this and five other leaching steps (Budde et al., 2019). However, only leaching step L2 (5.1 M $HNO_3$; 20 °C; 5 days), which mainly dissolved metal and sulfides and released the majority of the Mo and Ru of the bulk meteorite (~53% and ~69%), contained enough Mo and Ru to precisely analyze the isotope composition of both elements on the same solution. We therefore could only obtain Ru isotopic data for leaching step L2. The Mo isotopic composition of this leachate indicates a large *s*-process deficit (Budde et al., 2019), and so this sample is well suited for more precisely defining the Mo-Ru *s*-process correlation among NC meteorites.

## 2.2. Chemical purification of Ru and Mo

Ruthenium was separated by a three-stage ion exchange chromatography method described in Hopp and Kleine (2018). After conversion into chloride form, the samples were re-dissolved in 5 to 10 ml 0.2 M HCl and loaded onto cation exchange columns filled with 10 ml pre-cleaned BioRad AG 50W-X8 (100-200 mesh) resin. From these columns, the bulk of the HSE was eluted in a total volume of 14 ml 0.2 M HCl, while the major elements (i.e., Fe and Ni) remain adsorbed on the resin (e.g., Fischer-Gödde et al., 2010; Hopp et al., 2016). The HSE fractions were dried on a hotplate at 110°C, re-dissolved in 1 ml 0.2 M HCl-10% $Br_2$, and loaded onto columns filled with 0.25 ml BioRad Macro-Prep DEAE resin. Ruthenium was eluted in 6 ml of the same solution, whereas Pd is quantitatively adsorbed onto the resin as a bromine complex. The Ru fractions were dried and re-dissolved three times using 10 ml 1M HF. To remove remaining Zr and Mo, the Ru fractions were dissolved in 7 ml 1M HF and loaded onto anion exchange columns filled with 2 ml of pre-cleaned BioRad AG 1-X8 (100-200 mesh) resin. Ruthenium was eluted in 14 ml 1 M HF, whereas Zr and Mo are adsorbed onto the resin. The final Ru fractions were dried and re-dissolved in 0.5 ml 0.28M $HNO_3$ at 100°C. This procedure result in pure Ru fractions with Mo/Ru < 0.0005, Zr/Ru < 0.004, and Pd/Ru < 0.0005, which allow accurate correction of interferences on $^{98}Ru$, $^{100}Ru$, $^{102}Ru$ and $^{104}Ru$ (Fischer-Gödde et al., 2015). The overall Ru yield of the chemical separation varied between ~50 and ~90 %. Finally, the Ru blank for the whole procedure is 49 ± 33 pg Ru (1 s.d.; n=5) and insignificant, given that more than 70 ng Ru was analyzed for each sample.

The chemical separation of Mo was accomplished by ion exchange chromatography following the analytical protocol described by Budde et al. (2018, 2019), as summarized below. Molybdenum was separated from most of the sample matrix by loading the samples in 75 ml 0.5 M HCl-0.5 M HF onto columns filled with 4 ml of pre-cleaned Bio-Rad AG1-X8 anion exchange resin (200-400 mesh). After additional rinses with 10 ml 0.5 M HCl-0.5 M HF and 15 ml 6 M HCl-1 M HF (HFSE cut), Mo was collected with 10 ml of 3 M $HNO_3$. Samples of ~1 g were dissolved in 150 ml 0.5 M HCl-0.5 M HF and then processed consecutively in 2 splits over the same column (HFSE and the Mo cuts from the different splits were re-combined afterwards). Minor amounts of Mo (~15%) typically eluted together with the HFSE were recovered by loading the samples in 6 ml 0.6 M HF-0.4% $H_2O_2$ onto Bio-Rad Poly-Prep columns containing 1 ml of pre-cleaned Bio-Rad AG1-X8 anion exchange resin (200-400 mesh). After rinsing with different HCl-HF mixtures, Mo was eluted with 5 ml of 3 M $HNO_3$. The two Mo cuts were then combined and Mo concentrations were determined on small aliquots using a Thermo Scientific *XSeries 2* quadrupole ICP-MS. Finally, Mo was further purified using columns filled with 1 ml of pre-cleaned Eichrom TRU resin (100-150 μm). The samples were loaded in 1 ml 1M HCl and, after rinsing with 6 ml 1 M HCl, Mo eluted with 6 ml 0.1 M HCl. This chemistry was repeated once, but using 7 M $HNO_3$ and 0.1 M $HNO_3$ instead of 1 M HCl and 0.1 M HCl, respectively. All Mo cuts were evaporated with $HNO_3$ and inverse *aqua regia.* The Mo yields were typically ~75%, and the total procedural blanks were ~2-4 ng and thus negligible. The final Mo cuts of the samples typically had Ru/Mo and Zr/Mo of < 0.0001 and thus allow accurate correction of isobaric interferences (Budde et al., 2016).

*2.3. Mass spectrometry and data reduction*

The Ru isotope analyses were performed using the Neptune *Plus* MC-ICPMS at the Institut für Planetologie in Münster and followed the measurement protocol described in Fischer-Gödde et al. (2015). All samples and standards were dissolved in 0.28 M HNO$_3$ and the solutions were introduced into the mass spectrometer using a CETAC Aridus II desolvating system combined with a Savillex C-Flow nebulizer with a 50 μl/min uptake rate. During each session the formation of oxides, measured as CeO/Ce, was reduced to <1% through the addition of N$_2$ to the sample gas. Standards and sample solutions were measured at concentrations of ~100 ppb Ru using conventional Ni H cones. Ion beams were simultaneously collected in static mode for all seven stable Ru isotopes ($^{96}$Ru, $^{98}$Ru, $^{99}$Ru, $^{100}$Ru, $^{101}$Ru, $^{102}$Ru, $^{104}$Ru) together with $^{97}$Mo and $^{105}$Pd as interference monitors. Ion beams at the Ru masses were measured using Faraday cups connected to $10^{11}$ Ω feedback resistors, whereas ion beams at $^{97}$Mo and $^{105}$Pd were collected using Faraday cups connected to $10^{12}$ Ω resistors. Prior to each measurement, baselines were measured on peak with 40 × 8.4 s integrations on a solution blank. Each sample or standard measurement consisted of 100 × 8.4 s integrations. The sample analyses were bracketed by measurements of an in-house Ru solution standard (Alfa Aesar Ru). Mass bias was corrected by internal normalization to $^{99}$Ru/$^{101}$Ru = 0.7450754 (Chen et al., 2010) using the exponential law. The Ru isotope data are reported as the parts-per-10,000 deviations ($\varepsilon^i$Ru = [($^i$Ru/$^{101}$Ru) / ($^i$Ru/$^{101}$Ru)$_{standard}$ − 1] × $10^4$) from the average bracketing standard runs. The external reproducibility of the Ru isotope measurements (±0.13 for $\varepsilon^{100}$Ru; 2 s.d.) was evaluated using seven digestions of subsamples of a ~100 g powder of the Allende CV3 chondrite ('Allende MS-A') and three digestions of different pieces of the IIIAB iron meteorite Henbury (Supplementary material; Table S1).

Molybdenum isotope measurements were also performed using the Neptune *Plus* MC-ICPMS at Münster and followed the measurement protocol described in Budde et al. (2018, 2019). The samples were introduced into the mass spectrometer in 0.5M HNO$_3$-0.01M HF using a CETAC Aridus II desolvating system combined with a Savillex C-Flow nebulizer with a 50 μl/min uptake rate. Standards and sample solutions were measured at concentrations of ~100 ppb Mo using conventional Ni H cones and each measurement consisted of 40 × 8.4 s baseline integrations and 100 × 8.4 s isotope ratio measurements. Mass bias was corrected by internal normalization to $^{98}$Mo/$^{96}$Mo = 1.453173 using the exponential law. Isobaric interferences of Zr and Ru on Mo masses were monitored by simultaneous measurements of $^{91}$Zr and $^{99}$Ru using Faraday cups connected to $10^{12}$ Ω feedback resistors. The Mo isotope data are reported as the

parts-per-10,000 deviation ($\varepsilon^i Mo = [(^i Mo/^{96}Mo) / (^i Mo/^{96}Mo)_{standard} - 1] \times 10^4$) from the average bracketing standard runs. The external reproducibility of the Mo isotope measurements ranges from ±0.15 for $\varepsilon^{97}Mo$ to ±0.35 for $\varepsilon^{92}Mo$ (2 s.d.; see Budde et al., 2019), as defined by repeated analyses of the BHVO-2 rock standard (n=40) that was processed and measured together with each set of samples.

## 3. Results

The Mo isotope data are summarized in Table 1 and shown in a diagram of $\varepsilon^{95}Mo$ *vs.* $\varepsilon^{94}Mo$ (Fig. 1). In this plot, non-carbonaceous (NC) and carbonaceous (CC) meteorites plot on two parallel lines (the NC- and CC-lines; Budde et al., 2016), which predominantly reflect *s*-process variations within each reservoir (Budde et al., 2016, 2019; Poole et al., 2017; Worsham et al., 2017; Bermingham et al., 2018). Most of the samples from this study plot on the NC-line (Fig. 1); these samples display a similar range of Mo isotopic compositions as observed previously for NC meteorites, but include samples with the largest anomalies (NWA 6112 and NWA 1058) measured among NC meteorites so far ($\varepsilon^{94}Mo$ up to ~1.5). The two ureilites as well as brachinites and (brachinite-like) achondrites (NWA 6112, NWA 5363, NWA 1058) display variable *s*-process deficits and plot along the NC-line, indicating nucleosynthetic isotope heterogeneities within a given group of meteorites. Finally, the aubrite (Norton County), the enstatite achondrite (NWA 2526), and the winonaites (HaH 193, Winona) have the smallest Mo isotope anomalies of this sample set ($\varepsilon^{94}Mo$ from ~0.25 to ~0.60), which is similar to the range of Mo isotope anomalies observed for enstatite and ordinary chondrites (Render et al., 2017).

Two of the ungrouped primitive achondrites (NWA 8548, NWA 6926) plot on the CC-line and , therefore, are genetically linked to carbonaceous meteorites (Fig. 1). This is consistent with the O isotopic signatures of NWA 8548 suggesting a connection to CR chondrites, and with the O and Ti isotope signatures of the paired NWA 6926/6704 meteorites (e.g., Greenwood et al., 2017; Hibiya et al., 2019). However, the Mo isotope anomalies of NWA 8548 (e.g., $\varepsilon^{94}Mo$ = 1.53±0.10) are distinct from those of CR chondrites, which are characterized by a much larger *s*-process Mo deficit (e.g., $\varepsilon^{94}Mo$ = 3.11±0.15) (Budde et al., 2018).

The Ru isotopic data are summarized in Table 2 and Fig. 2. The samples of this study display variable $\varepsilon^{100}Ru$ from ~ +0.24 to -1.14, including the first bulk meteorites with positive $\varepsilon^{100}Ru$ (i.e., NWA 3151, NWA 4882). As for Mo isotopes, aubrites, the enstatite achondrite, and the winonaite show small negative $\varepsilon^{100}Ru$ (-0.06 to -0.08), similar to that of enstatite chondrites (Fischer-Gödde and Kleine, 2017). By contrast, ureilites, mesosiderites, acapulcoite-lodranites, and two ungrouped NC achondrites have Ru isotope anomalies similar to those of ordinary chondrites and NC iron meteorites (Fischer-Gödde et al., 2015). Two ungrouped achondrites (NWA 8548, NWA 6926), which on the basis of Mo isotopes belong to the CC suite of meteorites, have distinctively larger anomalies of $\varepsilon^{100}$Ru ≈ –1, similar to those observed for CC iron meteorites and bulk carbonaceous chondrites (Fischer-Gödde and Kleine, 2017; Worsham et al., 2019).

Finally, the L2 leachate of the ordinary chondrite NWA 2458 has the largest Ru isotope anomaly measured in this study. The $\varepsilon^{100}$Ru anomaly of –1.7 is consistent with a deficit in *s*-process nuclides, as observed in the Mo isotopic composition of this sample (Budde et al., 2019; Table 1). As will be discussed in more detail below, the leachate plots on the *s*-process Mo-Ru correlation line defined by bulk NC meteorites, indicating that incongruent dissolution of Mo and Ru carrier phases did not occur during this leaching step.

## 4. Discussion

Figure 3 displays the results of this study together with literature data for chondrites and iron meteorites in a plot of $\varepsilon^{100}$Ru *vs*. $\varepsilon^{94}$Mo. Combined, this is the largest and most comprehensive Mo-Ru isotope data set available to date. We note that several carbonaceous chondrites (e.g., CI, CM) are not included in this plot, because for these the Mo and Ru isotope data were obtained on different samples. As carbonaceous chondrites are isotopically heterogeneous for Mo and Ru (Budde et al., 2016; Fischer-Gödde and Kleine, 2017), this may result in deviations from the bulk Mo-Ru isotopic composition for these chondrites. Therefore, only carbonaceous chondrites for which Mo and Ru isotopes were measured on the same sample split are shown in Fig. 3.

The comprehensive data set shown in Fig. 3 has important implications for assessing the genetic characteristics of Earth's late-stage accretionary assemblage. Contrary to prior studies, we find that several NC meteorites, and most CC meteorites, deviate from the Mo-Ru cosmic correlation. Consequently, before the Mo-Ru isotopic data can be used to reconstruct Earth's accretion history (sect. 4.3), we will first evaluate the origin of these deviations from the Mo-Ru correlation (sect. 4.1), and precisely define the position of the BSE with respect to the Mo-Ru correlation line defined solely by NC meteorites (sect. 4.2).

**Table 1:** Molybdenum isotope data of primitive achondrites, aubrites, and mesosiderites.

| Sample | Weight (g) | Mo (µg/g)[a] | N[b] | ε$^{92}$Mo[c] | ε$^{94}$Mo[c] | ε$^{95}$Mo[c] | ε$^{97}$Mo[c] | ε$^{100}$Mo[c] |
|---|---|---|---|---|---|---|---|---|
| *Aubrites* | | | | | | | | |
| Norton County (metal)[d] | 0.501 | 2.85 | 6 | 0.57 ± 0.14 | 0.47 ± 0.05 | 0.26 ± 0.06 | 0.19 ± 0.03 | 0.17 ± 0.11 |
| Peña Blanca Spring[d] | 10.50 | 0.01 | 1 | 0.50 ± 0.35 | 0.58 ± 0.22 | 0.16 ± 0.15 | 0.07 ± 0.15 | 0.16 ± 0.22 |
| Wtd. average (± 2σ)[d] | | | | 0.56 ± 0.13 | 0.48 ± 0.05 | 0.25 ± 0.06 | 0.19 ± 0.03 | 0.16 ± 0.10 |
| *Mesosiderites* | | | | | | | | |
| Acfer 063[d] | 0.399 | 2.37 | 6 | 1.23 ± 0.21 | 1.05 ± 0.12 | 0.46 ± 0.07 | 0.26 ± 0.04 | 0.21 ± 0.09 |
| Ilafegh 002[d] | 0.336 | 2.43 | 7 | 1.16 ± 0.23 | 1.01 ± 0.16 | 0.45 ± 0.12 | 0.26 ± 0.08 | 0.18 ± 0.05 |
| NWA 2538[d] | 0.328 | 1.81 | 5 | 1.27 ± 0.34 | 1.04 ± 0.17 | 0.49 ± 0.12 | 0.24 ± 0.06 | 0.23 ± 0.14 |
| Wtd. average (± 2σ)[d] | | | (3) | 1.21 ± 0.14 | 1.04 ± 0.08 | 0.46 ± 0.05 | 0.25 ± 0.03 | 0.19 ± 0.04 |
| *Winonaites* | | | | | | | | |
| HaH 193 | 0.803 | 0.43 | 3 | 0.38 ± 0.35 | 0.29 ± 0.22 | 0.14 ± 0.15 | 0.09 ± 0.15 | 0.04 ± 0.22 |
| HaH 193[e] | – | – | 1 | 0.40 ± 0.90 | 0.25 ± 0.30 | 0.08 ± 0.15 | 0.05 ± 0.03 | 0.05 ± 0.22 |
| Winona[e] | – | – | 1 | 0.20 ± 0.90 | 0.18 ± 0.30 | 0.05 ± 0.15 | -0.01 ± 0.03 | 0.01 ± 0.22 |
| Wtd. average (± 2σ) | | | (3) | 0.36 ± 0.31 | 0.25 ± 0.15 | 0.09 ± 0.09 | 0.02 ± 0.02 | 0.03 ± 0.13 |
| *Acapulcoite-Lodranites* | | | | | | | | |
| DHO 125[d] | 0.535 | 0.91 | 5 | 1.01 ± 0.24 | 0.94 ± 0.12 | 0.41 ± 0.07 | 0.26 ± 0.08 | 0.26 ± 0.11 |
| GRA 95209 (metal)[e] | – | – | 1 | 1.48 ± 0.90 | 1.10 ± 0.30 | 0.48 ± 0.15 | 0.21 ± 0.03 | 0.24 ± 0.22 |
| MET 01195,42 | 0.524 | 1.14 | 5 | 1.05 ± 0.13 | 0.89 ± 0.09 | 0.49 ± 0.03 | 0.25 ± 0.05 | 0.22 ± 0.09 |
| Wtd. average (± 2σ) | | | (3) | 1.05 ± 0.11 | 0.92 ± 0.07 | 0.48 ± 0.03 | 0.23 ± 0.02 | 0.24 ± 0.07 |
| *Ureilites* | | | | | | | | |
| NWA 7630[d] | 0.798 | 0.53 | 3 | 1.06 ± 0.35 | 1.01 ± 0.22 | 0.44 ± 0.15 | 0.26 ± 0.15 | 0.26 ± 0.22 |
| Dho 1519[d] | 0.820 | 0.76 | 4 | 0.64 ± 0.19 | 0.61 ± 0.14 | 0.32 ± 0.10 | 0.22 ± 0.08 | 0.03 ± 0.12 |
| *Brachinites* | | | | | | | | |
| NWA 3151[d] | 0.531 | 0.47 | 3 | 1.29 ± 0.35 | 1.14 ± 0.22 | 0.61 ± 0.15 | 0.36 ± 0.15 | 0.38 ± 0.22 |
| NWA 4882[d] | 1.102 | 0.38 | 4 | 1.34 ± 0.16 | 1.10 ± 0.21 | 0.56 ± 0.11 | 0.32 ± 0.04 | 0.31 ± 0.10 |
| NWA 10637 | 1.029 | 1.05 | 7 | 1.35 ± 0.14 | 1.22 ± 0.09 | 0.58 ± 0.07 | 0.34 ± 0.02 | 0.29 ± 0.05 |
| *Prim. enstatite achondrite* | | | | | | | | |
| NWA 2526 | 0.519 | 0.60 | 4 | 0.85 ± 0.38 | 0.63 ± 0.20 | 0.41 ± 0.19 | 0.21 ± 0.10 | 0.03 ± 0.13 |
| Replicate | 1.027 | 0.54 | 4 | 0.76 ± 0.15 | 0.59 ± 0.17 | 0.37 ± 0.17 | 0.21 ± 0.06 | 0.01 ± 0.16 |
| Wtd. average (± 2σ) | | | (2) | 0.77 ± 0.14 | 0.60 ± 0.13 | 0.39 ± 0.13 | 0.21 ± 0.05 | 0.02 ± 0.10 |
| *Ungrouped achondrites* | | | | | | | | |
| NWA 6112[f] | 0.517 | 0.89 | 3 | 1.83 ± 0.35 | 1.55 ± 0.22 | 0.79 ± 0.15 | 0.49 ± 0.15 | 0.51 ± 0.22 |
| NWA 5400/5363[g] | 0.513 | 0.57 | 3 | 0.81 ± 0.35 | 0.66 ± 0.22 | 0.31 ± 0.15 | 0.18 ± 0.15 | 0.14 ± 0.22 |
| NWA 1058[d] | 0.514 | 1.41 | 6 | 1.61 ± 0.12 | 1.31 ± 0.11 | 0.68 ± 0.09 | 0.38 ± 0.10 | 0.40 ± 0.08 |
| NWA 11048[h] | 0.513 | 1.34 | 5 | 0.53 ± 0.19 | 0.56 ± 0.11 | 0.28 ± 0.12 | 0.15 ± 0.07 | 0.09 ± 0.05 |
| NWA 8548 | 1.035 | 0.87 | 5 | 2.24 ± 0.10 | 1.53 ± 0.10 | 1.27 ± 0.07 | 0.64 ± 0.05 | 0.71 ± 0.10 |
| NWA 6926 | 1.064 | 0.91 | 8 | 1.99 ± 0.12 | 1.48 ± 0.12 | 1.14 ± 0.07 | 0.59 ± 0.06 | 0.62 ± 0.06 |
| *Ordinary chondrite leachate* | | | | | | | | |
| NWA 2458 L2[d] | – | – | 6 | 3.90 ± 0.20 | 3.18 ± 0.08 | 1.87 ± 0.09 | 0.98 ± 0.05 | 1.28 ± 0.07 |

[a] Molybdenum concentrations were determined by quadrupole ICP-MS, which have an uncertainty of ~5%.
[b] Number of analyses of the sample solution.
[c] Molybdenum isotope data are internally normalized to $^{98}$Mo/$^{96}$Mo = 1.453173. Given uncertainties represent the external reproducibility (2 s.d.) obtained from repeated analyses of BHVO-2 (Budde et al., 2019) or 95% confidence intervals (95% c.i.) for samples with N>3.
[d] Molybdenum isotope data from Budde et al. (2019).
[e] Molybdenum isotope data from Worsham et al. (2017).
[f] NWA 6112 is petrologically similar to brachinites, however, based on the Fa composition slightly outside the range of brachinites, hence, classified as ungrouped achondrite. Recently, Hasegawa et al. (2019) re-investigated oxygen isotopes, petrography, mineralogy, and olivine fabric of NWA 6112 and concluded that the sample belongs to the brachinite clan meteorites.

[g] Paired with NWA 5363. Classified as brachinite-like meteorite (Day et al., 2012).
[h] NWA 11048 is classified as acapulcoite based on mineral chemistry and mean grain size of silicates. However, the nucleosynthetic Mo isotope anomaly of this sample is different compared to acapulcoite-lodranites.

**Table 2:** Ruthenium isotope data of primitive achondrites, aubrites, and mesosiderites.

| Sample | Weight (g) | N[a] | $\varepsilon^{96}Ru^b$ | $\varepsilon^{98}Ru^b$ | $\varepsilon^{100}Ru^b$ | $\varepsilon^{102}Ru^b$ | $\varepsilon^{104}Ru^b$ |
|---|---|---|---|---|---|---|---|
| *Aubrites* | | | | | | | |
| Norton County (metal) | 0.883 | 10 | -0.15 ± 0.22 | 0.08 ± 0.30 | -0.06 ± 0.03 | -0.11 ± 0.06 | -0.06 ± 0.17 |
| *Mesosiderites* | | | | | | | |
| Acfer 063 | 0.552 | 10 | 0.82 ± 0.70 | 0.70 ± 0.93 | -0.43 ± 0.03 | -0.18 ± 0.09 | 0.32 ± 0.17 |
| Ilafegh 002 | 0.792 | 11 | 0.93 ± 0.66 | 0.66 ± 0.53 | -0.42 ± 0.02 | -0.11 ± 0.09 | 0.34 ± 0.31 |
| Wtd. average (± 2σ) | | (2) | 0.88 ± 0.48 | 0.67 ± 0.46 | -0.42 ± 0.02 | -0.15 ± 0.06 | 0.32 ± 0.15 |
| *Winonaites* | | | | | | | |
| Hammadah al Hamra 193 | 0.803 | 4 | 0.52 ± 0.35 | 0.39 ± 0.71 | -0.06 ± 0.10 | 0.07 ± 0.13 | 0.40 ± 0.23 |
| *Acapulcoite-Lodranites* | | | | | | | |
| Dhofar 125 | 0.504 | 3 | 0.94 ± 0.55 | 0.49 ± 0.41 | -0.31 ± 0.08 | -0.04 ± 0.03 | 0.54 ± 0.18 |
| MET 01195,42 | 0.509 | 7 | 0.35 ± 0.52 | 0.73 ± 0.43 | -0.38 ± 0.11 | -0.04 ± 0.09 | 0.68 ± 0.19 |
| GRA 95209,274 | 0.308 | 7 | 0.90 ± 0.29 | 0.77 ± 0.26 | -0.33 ± 0.10 | -0.10 ± 0.12 | 0.51 ± 0.26 |
| NWA 7474 | 0.920 | 6 | 0.29 ± 0.31 | 0.29 ± 0.56 | -0.37 ± 0.11 | -0.14 ± 0.09 | 0.22 ± 0.08 |
| Wtd. average (± 2σ) | | (4) | 0.62 ± 0.56 | 0.65 ± 0.18 | -0.34 ± 0.05 | -0.05 ± 0.06 | 0.34 ± 0.32 |
| *Ureilites* | | | | | | | |
| NWA7630 | 0.729 | 3 | 0.78 ± 0.45 | 0.13 ± 0.52 | -0.31 ± 0.13 | -0.17 ± 0.15 | 0.04 ± 0.31 |
| DHO1519 | 0.530 | 1 | -0.87 ± 0.45 | -0.26 ± 0.52 | -0.23 ± 0.13 | -0.31 ± 0.15 | -0.12 ± 0.31 |
| *Brachinites* | | | | | | | |
| NWA 3151 | 1.004 | 4 | 1.06 ± 0.53 | 0.71 ± 0.87 | 0.16 ± 0.11 | 0.15 ± 0.21 | 0.57 ± 0.44 |
| Replicate | 1.003 | 2 | 0.97 ± 0.45 | 0.51 ± 0.52 | 0.26 ± 0.13 | 0.06 ± 0.15 | 0.22 ± 0.31 |
| Replicate | 1.001 | 2 | 0.76 ± 0.45 | 0.36 ± 0.52 | 0.36 ± 0.13 | 0.22 ± 0.15 | 0.33 ± 0.31 |
| Wtd. average (± 2σ)[c] | | (3) | 0.92 ± 0.27 | 0.48 ± 0.33 | 0.25 ± 0.07 | 0.14 ± 0.10 | 0.33 ± 0.20 |
| NWA 4882 | 1.084 | 3 | 1.05 ± 0.45 | 0.40 ± 0.52 | 0.11 ± 0.13 | 0.03 ± 0.15 | 0.17 ± 0.31 |
| NWA 10637 | 0.974 | 8 | 0.54 ± 0.12 | 0.12 ± 0.32 | -0.14 ± 0.06 | 0.06 ± 0.07 | 0.63 ± 0.15 |
| Replicate | 0.741 | 8 | 0.33 ± 0.27 | 0.32 ± 0.19 | -0.05 ± 0.08 | 0.01 ± 0.04 | 0.20 ± 0.13 |
| Wtd. average (± 2σ) | | (2) | 0.51 ± 0.11 | 0.27 ± 0.16 | -0.11 ± 0.05 | 0.02 ± 0.04 | 0.40 ± 0.30 |
| *Prim. enstatite achondrite* | | | | | | | |
| NWA 2526 | 1.001 | 2 | 0.56 ± 0.45 | 0.56 ± 0.52 | -0.08 ± 0.13 | 0.11 ± 0.15 | 0.51 ± 0.31 |
| *Ungrouped achondrites* | | | | | | | |
| NWA 6112[c] | 1.008 | 8 | 0.53 ± 0.22 | 0.45 ± 0.30 | -0.46 ± 0.07 | -0.14 ± 0.07 | 0.23 ± 0.09 |
| NWA 5363/5400[d] | - | - | 0.58 ± 0.44 | 0.06 ± 0.55 | -0.34 ± 0.13 | -0.19 ± 0.14 | 0.08 ± 0.35 |
| NWA 1058 | 1.049 | 6 | 1.09 ± 0.85 | 0.75 ± 0.76 | -0.40 ± 0.07 | -0.09 ± 0.11 | 0.50 ± 0.19 |
| NWA 8548 | 0.986 | 6 | 0.33 ± 0.42 | -0.30 ± 0.35 | -1.14 ± 0.10 | -0.34 ± 0.17 | 0.17 ± 0.34 |
| NWA 6926 | 0.652 | 4 | 0.02 ± 0.37 | -0.24 ± 0.66 | -1.04 ± 0.14 | -0.39 ± 0.18 | -0.22 ± 0.08 |
| *Ordinary chondrite leachate* | | | | | | | |
| NWA 2458 L2 | | 6 | 1.20 ± 0.34 | 1.19 ± 0.32 | -1.71 ± 0.03 | -0.56 ± 0.05 | 0.88 ± 0.12 |

[a] Number of analyses of the sample solution.
[b] Ruthenium isotope ratios are internally normalized to $^{99}Ru/^{101}Ru$ = 0.7450754 using the exponential law and are reported relative to the Alfa Aesar bracketing data. Given uncertainties represent the external reproducibility (2 s.d.) reported in Table S1 or 95% confidence interval (95% c.i.) for samples with N>3.
[c] NWA 6112 is petrologically similar to brachinites, however, based on the Fa composition slightly outside the range of brachinite and hence classified as ungrouped achondrite. Recently, Hasegawa et al. (2019) re-investigated oxygen isotopes, petrography, mineralogy, and olivine fabric of NWA 6112 and concluded that the sample belongs to the brachinite clan meteorites.
[d] Paired with NWA 5400. Classified as brachinite-like meteorite (Day et al., 2012). Data from Burkhardt et al. (2017).

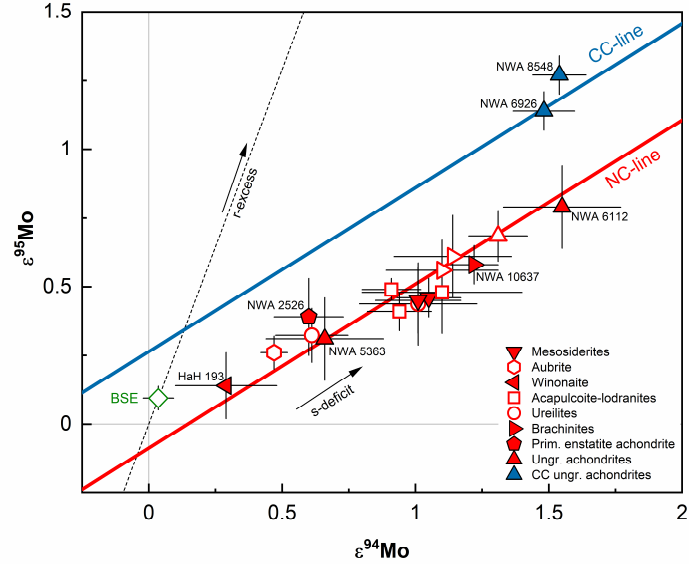

**Fig. 1:** Diagram of $\varepsilon^{95}$Mo *vs.* $\varepsilon^{94}$Mo for samples of this study. In this plot, the Mo isotope dichotomy between carbonaceous (CC, blue) and non-carbonaceous (NC, red) materials is defined by two parallel lines with identical slopes (Budde et al., 2019). Samples from both groups have variable *s*-process deficits relative to the bulk silicate Earth (BSE) and plot along each line. The offset between the lines reflects an approximately constant *r*-excess in the CC reservoir. Two of the ungrouped primitive achondrites (NWA 8548, NWA 6926) plot on to the CC-line, whereas all other samples of this study plot on the NC-line. Open symbols represent samples for which Mo isotopic data are reported in Budde et al. (2019), and which were investigated for Ru isotopes in this study. For samples shown with closed symbols, both Mo and Ru isotopic data were obtained in this study. Slopes of NC-, CC-, and *r*-process mixing lines as well as the composition of the BSE are from Budde et al. (2019).

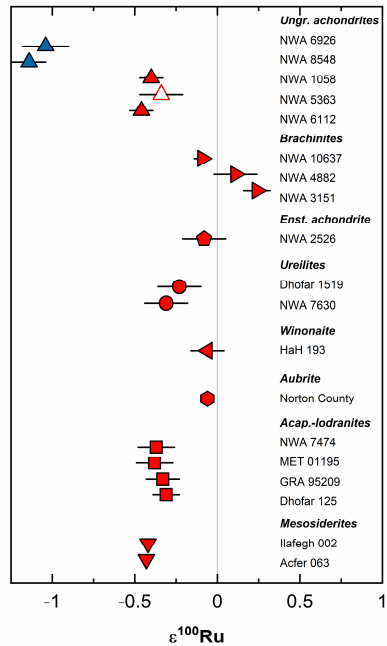

**Fig. 2:** Ruthenium isotope data for samples of this study. CC meteorites are shown with blue, NC meteorites with red symbols. Brachinites and brachinite-like ungrouped achondrites have variable Ru isotope anomalies, and are the first group of meteorites with positive $\varepsilon^{100}$Ru. The Ru isotopic data for NWA 5363 (open symbol) is from Burkhardt et al. (2017).

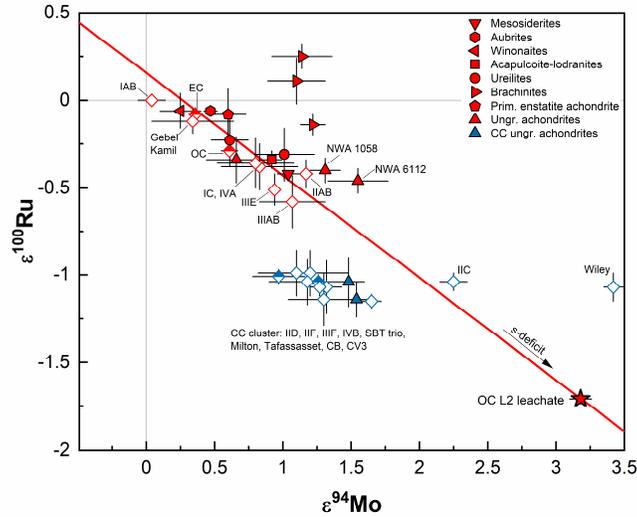

**Fig. 3:** Diagram of $\varepsilon^{100}$Ru *vs.* $\varepsilon^{94}$Mo for samples of this study together with literature data for chondrites (half-open diamonds) and iron meteorites (open diamonds). Except for brachinites, NC meteorites (red symbols) show a broad correlation between $\varepsilon^{94}$Mo and $\varepsilon^{100}$Ru, whereas CC meteorites (blue symbols) have indistinguishable $\varepsilon^{100}$Ru of ~ -1, with most plotting in a cluster at $\varepsilon^{94}$Mo ≈ 1.3. The red solid line is the *s*-process mixing line defined by NC meteorites as described in sect. 4.2 and Fig. 4. Literature data from Fischer-Gödde et al. (2015), Render et al. (2017), Fischer-Gödde and Kleine (2017), Burkhardt et al. (2017), Poole et al. (2017), Bermingham et al. (2018), Budde et al. (2019), Worsham et al. (2019), and Hilton et al. (2019).

*4.1. Decoupled Mo-Ru isotope systematics in partially differentiated meteorites*

Consistent with prior studies, most NC meteorites from this study display correlated Mo and Ru isotope anomalies. This includes winonaites, aubrites, acapulcoites-lodranites, mesosiderites, ureilites, and the ungrouped achondrites NWA 5363/5400 (Fig. 3). By contrast, the three brachinites investigated in this study as well as the brachinite-like achondrites NWA 6112 (and perhaps NWA 1058) plot off the Mo-Ru correlation line defined by the other NC meteorites (Fig. 3). These samples have similar $\varepsilon^{94}$Mo, but distinct $\varepsilon^{100}$Ru; as such, their Mo and Ru isotope anomalies are decoupled and cannot be accounted for by the heterogeneous distribution of a common *s*-process carrier, unlike for all other NC meteorites investigated thus far. The two ureilites of this study also show hints for decoupled Mo-Ru isotope systematics, but plot within uncertainty of the Mo-Ru *s*-process correlation line.

Worsham et al. (2019) observed decoupled Mo and Ru isotope anomalies in carbonaceous iron meteorites, which despite of large Mo isotope variations have an approximately constant $\varepsilon^{100}$Ru of *ca*. –1. Worsham et al. (2019) interpreted this to reflect the preferential loss of *s*-process Mo isotopes from precursor dust by thermal processing under oxidizing conditions in the outer solar system (i.e., the formation location of CC meteorites). This raises the question of whether the decoupled Mo-Ru isotope systematics observed for brachinites are caused by similar processes. However, whereas in the CC reservoir the decoupling of Mo and Ru isotopes is observed for iron meteorites (Worsham et al., 2019) and probably also some chondrites (i.e., the CR chondrites; Budde et al., 2018), this is not true for the NC reservoir. Instead, among the NC meteorites, decoupled Mo and Ru isotope anomalies predominantly occur in partially differentiated meteorites, in particular the brachinites. By contrast, meteorites from parent bodies that either escaped large-scale melting and melt migration (i.e., chondrites, winonaites, acapulcoites-lodranites) or underwent complete differentiation (i.e., iron meteorites, aubrites, mesosiderites) plot close to the Mo-Ru *s*-process correlation line (Fig. 3). This observation suggests that the incomplete differentiation of the parent bodies of brachinites (and brachinite-like meteorites) may have been responsible for producing the intra-group heterogeneity and decoupled Mo-Ru isotope systematics.

Goderis et al. (2015) first proposed a planetary differentiation origin of nucleosynthetic isotope anomalies, on basis of nucleosynthetic Os isotope variations among ureilites. As iron meteorites and bulk chondrites have no resolved nucleosynthetic Os isotope anomalies (e.g., Yokoyama et al., 2007; Walker, 2012), Goderis et al. (2015) concluded that the Os isotope variations among ureilites do not reflect the heterogeneous distribution of presolar Os carriers in the solar nebula, but rather result from the heterogeneous dissolution of presolar carrier phases during localized melting events on the ureilite parent body. Subsequently, Budde et al. (2019) suggested that similar processes could account for the Mo *s*-process variability observed among ureilites. To account for decoupled Mo-Ru isotope systematics by this process requires disparate behavior of the Mo and Ru that was released from presolar carriers during partial melting and melt removal. Otherwise, the Mo and Ru isotope variations would follow the predicted *s*-process correlation line (Fig. 3). This may be the case for the two ureilites, but not for the brachinites, which plot off the Mo-Ru *s*-process correlation line.

Under oxidizing conditions in the solar nebula, Mo may form volatile oxides and could then separate from the more refractory Ru (Fegley and Palme, 1985). However, although differentiation of the brachinite parent body occurred under relatively oxidizing conditions (e.g., Keil, 2014), it is unclear if the different behavior of Mo and Ru in specific solar nebula environments is relevant during partial melting and melt migration on meteorite parent bodies. Instead, it seems more likely that the distinct behavior of Mo and Ru during melting and melt migration on the brachinite parent body played some role. Differentiation of the brachinite parent body probably involved the separation of Fe-S melts from solid, S-poor Fe metal (Day et al., 2012). If the Fe-S melt was isotopically anomalous -because it was enriched or depleted in certain presolar carriers (Goderis et al., 2015)- removal of this melt would have resulted in a complementary isotopic signature in the residue. The magnitude of this effect would then depend on a how a given element partitioned between the melt and the residue. As Ru is more compatible in solid Fe metal than Mo (Hayden et al., 2011), this process may have affected Ru and Mo isotopes differently and, therefore, may have resulted in the observed decoupling of Ru and Mo isotope anomalies in the brachinites and brachinite-like achondrites.

The observation of nucleosynthetic isotope anomalies generated during planetary differentiation raises the question of whether a similar process also occurred on fully differentiated meteorite parent bodies, such as those of the iron meteorites. As the iron meteorites plot along the expected *s*-process correlation line, this process, if it occurred, evidently did not result in significant incongruent redistribution of isotopically anomalous Mo and Ru in the iron meteorite parent bodies. Nevertheless, planetary processing of presolar carriers may be responsible for some of the scatter around the Mo-Ru correlation line, such as, for instance, the observation that the IIAB irons plot slightly to the right of this line (Fig. 3).

*4.2. Mo-Ru correlation in the NC reservoir*

Despite the decoupled Mo-Ru isotope systematics observed for brachinites, most of the NC meteorites plot along the predicted *s*-process Mo-Ru correlation line (Fig. 3). These samples include enstatite and ordinary chondrites, iron meteorites, and several meteorite groups for which combined Mo-Ru isotopic data are reported for the first time in this study (i.e., acapulcoites-lodranites, winonaites, aubrites, mesosiderites). Some of the ungrouped primitive achondrites investigated here also plot on, or close to, the Mo-Ru correlation, but these samples may derive from partially differentiated bodies akin to that of the brachinites. As such, their Mo-Ru isotope systematics may have been modified by parent body processes. To assess how including such samples in the regression affects the slope and intercept of the Mo-Ru correlation line for NC meteorites, we calculated regressions including and excluding these samples (Tables S2 and S3). In addition, for each set of bulk meteorites used in the regression, we also calculated the regressions including the ordinary chondrite leachate. Only the brachinites are excluded from all regressions, because, as noted above, their Mo-Ru isotope systematics have been strongly decoupled. The data of this study together with data from previous studies now provide combined Mo-Ru isotopic compositions of at least 11 distinct NC parent bodies (see Table S2 for a summary of the data used in the regressions). This allows defining the Mo-Ru correlation more precisely than in previous studies, and, importantly, to define the correlation based solely on NC meteorites. By contrast, prior studies have used both NC and CC meteorites in the regression (e.g., Bermingham et al., 2018). However, NC and CC meteorites are not expected to plot on a single *s*-process Mo-Ru correlation line, because CC meteorites are characterized by an approximately constant excess in *r*-process Mo nuclides relative to the NC group (Budde et al., 2016; Poole et al., 2017; Worsham et al. 2017; Budde et al., 2019). The *r*-process excess becomes apparent only when Mo isotopes with *p*-process contributions are considered (i.e., $^{92}$Mo, $^{94}$Mo), because for these two isotopes *s*- and *r*-process Mo isotope variations result in distinct Mo isotope patterns (Burkhardt et al., 2011). Consistent with this and as shown in Fig. 3, Worsham et al. (2019) demonstrated that in plots of $\varepsilon^{100}$Ru *vs*. $\varepsilon^{92,94}$Mo CC meteorites plot off the *s*-process Mo-Ru correlation line defined by NC meteorites (Fig. 4a). By contrast, in plots of $\varepsilon^{100}$Ru *vs*. $\varepsilon^{95,97,100}$Mo, CC and NC meteorites plot on single *s*-process Mo-Ru correlation lines because *s*- and *r*-process variations result in mixing lines with similar slopes (Fig. 4b). As a result, for inferring the nature of Earth's late-stage building blocks, the $\varepsilon^{100}$Ru *vs*. $\varepsilon^{92,94}$Mo correlations are most useful, because only those allow discriminating between the accretion of NC and CC materials to Earth.

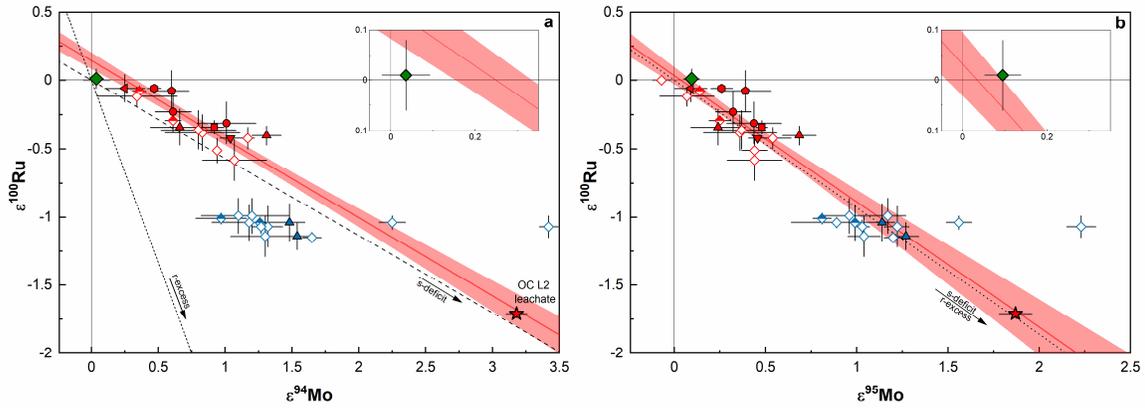

**Fig. 4:** Best-fits of the Mo-Ru isotope correlation for NC meteorites in **(a)** $\varepsilon^{100}$Ru *vs*. $\varepsilon^{94}$Mo **(b)** $\varepsilon^{100}$Ru *vs*. $\varepsilon^{95}$Mo space. The L2 acid leachate of the ordinary chondrite NWA 2458 (red star 'OC L2 leachate') falls exactly on the Mo-Ru correlation line defined by the NC meteorites. Red solid lines are linear regressions of the NC data for $\varepsilon^{100}$Ru *vs*. $\varepsilon^{94,95}$Mo including all NC meteorites (except brachinites) and the OC L2 leachate (shaded red area is 95% c.i. of best-fit). **(a)** In the $\varepsilon^{100}$Ru *vs*. $\varepsilon^{94}$Mo plot, the Mo-Ru correlation defines *x*-axis intercepts that are resolved from zero (Table S3) and CC meteorites do not plot on the correlation defined by NC meteorites. **(b)** In the $\varepsilon^{100}$Ru *vs*. $\varepsilon^{95}$Mo plot, the *x*-axis intercepts are not resolved from zero (Table S3) and the CC cluster falls on the correlation line defined by NC meteorites. Symbols and data of bulk meteorites as in Fig. 3 (and Table S2, S4). Data for BSE from Bermingham and Walker (2017) and Budde et al. (2019). Best-fits were calculated using the York method and the software suite OriginPro (OriginLab, Northampton, MA).

The results of the linear regression including all NC meteorites (except brachinites) and the OC leachate for $\varepsilon^{100}$Ru *vs*. $\varepsilon^{94}$Mo and $\varepsilon^{100}$Ru *vs*. $\varepsilon^{95}$Mo are plotted in Fig. 4 and summarized in

Table S3. For each set of samples, the regressions including the data for the ordinary chondrite leachate are consistent with those calculated without this data point, but, owing to the larger anomalies in the leachate, are a factor of about two more precise (Table S3). The results of the regressions do not change significantly whether or not ungrouped primitive achondrites are included in the regression (Table S3). Thus, any decoupling of Mo and Ru in these samples as a result of planetary processing is inconsequential for the overall definition of the NC line in Mo-Ru isotope space (with the exception of brachinites, which were excluded from the regressions). The $\varepsilon^{100}$Ru *vs.* $\varepsilon^{i}$Mo slopes calculated from the regressions are consistent with the *s*-process slopes predicted from presolar SiC data (Nicolussi et al., 1998; Arlandini et al., 1999; Savina et al., 2004; Stephan et al., 2019) indicating that the Mo and Ru isotope variations among the NC meteorites reflect the heterogeneous distribution of a common *s*-process carrier for Mo and Ru (Fig. 5). This conclusion is consistent with that of several prior studies (Dauphas et al., 2004; Burkhardt et al., 2011; Fischer-Gödde et al., 2015; Bermingham et al., 2018), with the important difference that with the new data the slopes of the correlations involving all Mo isotopes are precisely defined solely based on NC meteorites. Thus, only with the new data is it possible to reach firm conclusions about the nature of the correlated isotope anomalies among NC meteorites.

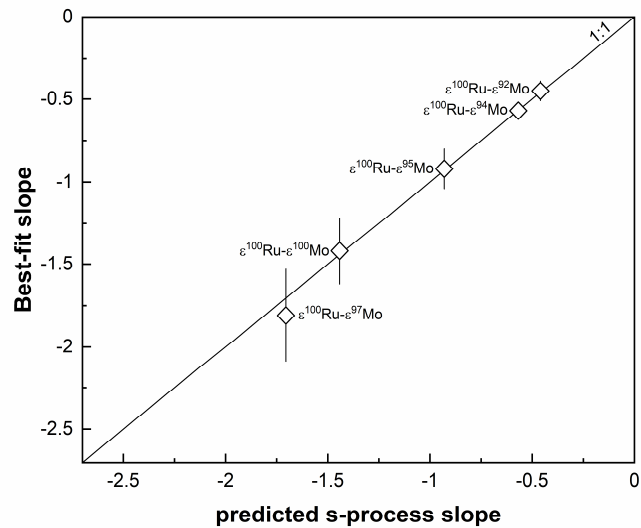

**Fig. 5:** Plot of the best-fit $\varepsilon^{100}$Ru *vs.* $\varepsilon^{i}$Mo slopes defined by the linear regressions of all NC meteorites (except brachinites) and the OC L2 leachate versus the predicted slopes of *s*-process mixing lines (Table S3). Predicted *s*-process slopes are based on data for presolar SiC grains (Arlandini et al., 1999; Savina et al., 2004; Stephan et al., 2019) and were calculated using equations described in Dauphas et al. (2004). Best-fits were calculated using the York method and the software suite OriginPro (OriginLab, Northampton, MA). Uncertainties are 95% c.i. as calculated by the linear regression (Table S3).

Contrary to previous studies, the Mo-Ru correlation line for NC meteorites reveals that the x-axis intercepts for the $\varepsilon^{100}$Ru vs. $\varepsilon^{94}$Mo and $\varepsilon^{100}$Ru vs. $\varepsilon^{92}$Mo correlations are resolved from zero, and from the BSE composition (Fig. 4a; Table S3). By contrast, for the $\varepsilon^{100}$Ru vs. $\varepsilon^{95,97,100}$Mo correlations the x-axis intercepts are not resolved, and the BSE plots on the correlation line (Fig. 4b; Table S3). Consequently, when isotope pairs that allow distinguishing between NC and CC meteorites are considered (i.e., $\varepsilon^{100}$Ru vs. $\varepsilon^{92,94}$Mo), then the BSE does *not* plot on the *s*-process Mo-Ru correlation line defined by NC meteorites. Moreover, both the BSE and the 'CC cluster' are offset from the Mo-Ru correlation towards *r*-process-enriched compositions (Fig. 4).

*4.3. Heterogeneous late-stage accretion of Earth*

The budgets of Mo and Ru in the BSE were established during different stages of Earth's accretion: Mo, as a moderately siderophile element, records the last 10–15 % of accretion, whereas Ru, as a highly siderophile element, predominantly records the late veneer (Dauphas, 2017; Bermingham et al., 2018). Thus, the observation that the BSE plots off the NC-line in $\varepsilon^{100}$Ru–$\varepsilon^{94}$Mo space indicates that the late stages of Earth's accretion were heterogeneous and involved genetically distinct building blocks.

The BSE may plot off the Mo-Ru correlation because Earth's late-stage building blocks derived from the NC reservoir but had distinct *s*-process anomalies. These objects would themselves plot on the Mo-Ru correlation, but because the proportions with which they contributed to the BSE's Mo and Ru varied, the BSE itself would plot away from the Mo-Ru correlation. In this case, the BSE would plot off the Mo-Ru correlation line in all plots of $\varepsilon^{100}$Ru vs. $\varepsilon^i$Mo, because all Mo isotope anomalies would be affected by the *s*-process variations in a similar manner. However, the BSE plots off the Mo-Ru correlation only for Mo isotopes for which *s*- and *r*-process variations result in distinct Mo isotope anomalies (i.e., the *p*-process nuclides $^{92}$Mo and $^{94}$Mo), but not for Mo isotopes that cannot distinguish between these variations (i.e., $\varepsilon^{95}$Mo, $\varepsilon^{97}$Mo, $\varepsilon^{100}$Mo) (Fig. 4). Consequently, the distinct Mo and Ru isotopic signatures of Earth's late-stage building blocks must, at least in part, reflect *r*-process (and possibly *p*-process) variations; such variations are characteristic for the isotopic difference between NC and CC meteorites, implying that Earth's late-stage building blocks included both NC and CC bodies.

This finding is consistent with results of Budde et al. (2019), who showed that in Mo isotope space the BSE plots between the NC- and CC-lines, indicating that the BSE's Mo derives from both the NC and the CC reservoir. Because the BSE's Mo records the last 10–15% of Earth's accretion (Dauphas, 2017), it predominantly represents the material accreted by the Moon-forming impactor and the late veneer. Thus, reproducing the BSE's Mo isotopic composition requires a CC heritage of either the Moon-forming impactor or the late veneer, or mixed NC-CC compositions for both (Budde et al., 2019). All three scenarios are consistent with the observation that the BSE plots off the NC-line in Mo-Ru isotope space. Nevertheless, the combined Mo-Ru data can be used to further evaluate these three possibilities.

It seems unlikely that the late veneer consisted entirely of CC material, because the BSE's $\varepsilon^{100}$Ru is most distinct from those of CC meteorites (Fischer-Gödde and Kleine, 2017; Worsham et al., 2019). Most CC meteorites have $\varepsilon^{100}$Ru values of around –1, but bulk carbonaceous chondrites display significant intra-group heterogeneity, most likely reflecting heterogeneities at the sampling scale (Fischer-Gödde and Kleine, 2017). Nevertheless, all carbonaceous chondrites analyzed to date have negative $\varepsilon^{100}$Ru and are distinct from BSE. Any contribution of such material to the late veneer would require that the late veneer also contained material with positive $\varepsilon^{100}$Ru, to counterbalance the negative $\varepsilon^{100}$Ru of known CC meteorites. The data of this study show that material with positive $\varepsilon^{100}$Ru exists (i.e., the brachinites). This opens up the possibility, but does not require, that the late veneer consisted of a heterogeneous mixture of materials with positive and negative $\varepsilon^{100}$Ru, including carbonaceous meteorites. Combined, these observations suggest that the late veneer had either a pure NC or a mixed NC-CC composition.

If the late veneer had a pure NC composition, then its Mo isotopic composition can be deduced from the $\varepsilon^{100}$Ru- $\varepsilon^{i}$Mo correlations. This is because the Ru isotopic composition of the BSE (i.e., $\varepsilon^{100}$Ru = 0) solely represents that of the late veneer, and so the corresponding $\varepsilon^{i}$Mo values are given by the x-axis intercepts of the $\varepsilon^{100}$Ru- $\varepsilon^{i}$Mo correlations for NC meteorites (Fig. 4). The Mo isotope pattern defined by these intercepts is similar to that of enstatite chondrites, albeit with slightly smaller anomalies (Fig. 6; Table S3). Further, in a diagram of $\varepsilon^{95}$Mo vs. $\varepsilon^{94}$Mo, this hypothetical late veneer composition plots on the NC-line, as expected from our starting assumption that the late veneer had a pure NC composition (Fig. 7). If the late veneer had this composition, then the BSE's Mo isotopic composition can only be reproduced by addition of *s*-process-enriched CC material through the Moon-forming giant impactor (Fig. 7) (Budde et al., 2019).

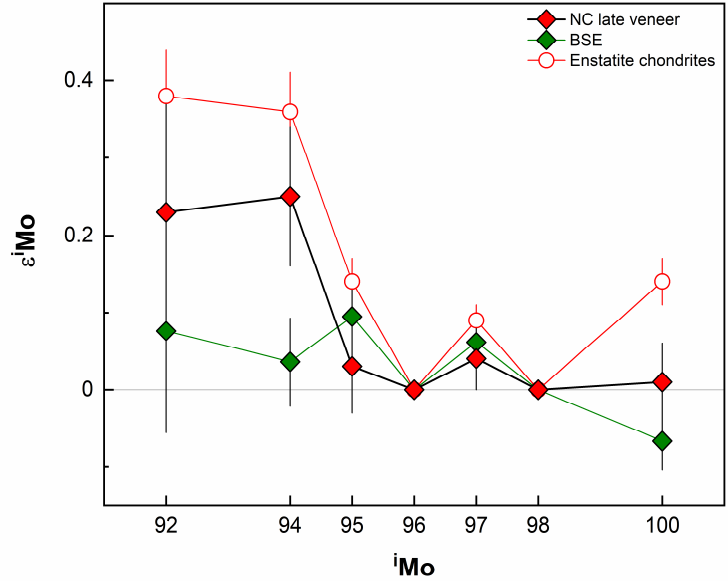

**Fig. 6:** ε$^i$Mo values of a hypothetical NC late veneer defined by the *x*-axis intercepts of the Mo-Ru correlation (Table S3). Results of the regressions using all NC meteorites (except brachinites) and the OC L2 leachate are shown. Data for enstatite chondrites and BSE are shown for comparison (Render et al., 2017; Budde et al., 2019). For *s*- and *r*-process nuclides ($^{95}$Mo, $^{97}$Mo, and $^{100}$Mo) the BSE and the hypothetical NC late veneer have indistinguishable Mo isotopic compositions. However, for the two *p*-process nuclides ($^{92}$Mo and $^{94}$Mo) the inferred composition of the hypothetical NC late veneer is significantly different from BSE. Uncertainties on the NC late veneer values are 95% c.i., as calculated by the linear regression (Table S3).

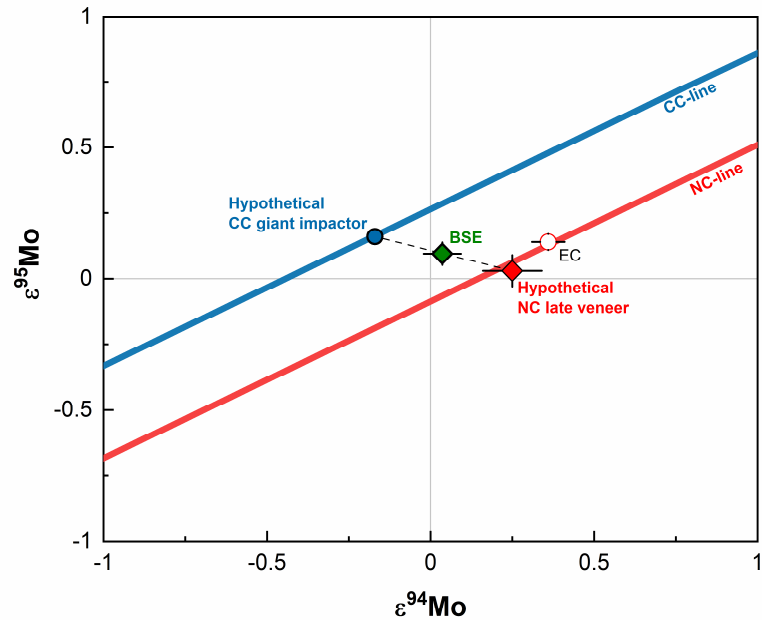

**Fig. 7:** Diagram of ε$^{95}$Mo *vs.* ε$^{94}$Mo illustrating the position of the BSE between the NC- and CC-lines (Budde et al., 2019). One possible endmember scenario for Earth's late-stage accretion involving a NC late veneer and a CC Moon-forming giant impactor is shown. The Mo isotopic composition of a hypothetical NC late veneer is given by the *x*-axis intercepts of the ε$^{100}$Ru *vs.* ε$^i$Mo correlations for NC meteorites (see Fig. 6) and falls exactly on the NC-line (red diamond). Based on this hypothetical late veneer composition and the BSE composition, the Moon-forming impactor was a CC body and had a small excess in *s*-process nuclides relative to the BSE (blue circle).

A strength of this model is that Earth's building material may have largely consisted of bodies with enstatite chondrite-like isotopic compositions, while CC material was only added during the final stages of accretion (Fig. 7). This provides a straightforward explanation for why the BSE's isotopic composition is similar to enstatite chondrites for elements that record Earth's entire accretion history (e.g., Ti, Cr, O), but deviates for Mo, which predominantly record the late stages of accretion (Budde et al., 2019).

It is noteworthy that in this model the *s*-process Mo isotope deficit (relative to the BSE) of the late veneer is small, and that, therefore, the complementary *s*-process excess in the Moon-forming impactor must also be small (Fig. 7). Thus, although the Moon-forming impactor and the late veneer would derive from distinct regions of the disk (namely the CC and NC reservoirs, respectively), their absolute Mo isotope anomalies (i.e., $\varepsilon^i$Mo values) were very similar to each other and to the (enstatite chondrite-like) proto-Earth (Budde et al., 2019). The reason for this apparent similarity (but genetic difference) remains unclear at present.

Another possibility is that the impactor and the late veneer had mixed NC-CC compositions. We emphasize that, as already pointed out by Budde et al. (2019), a mixed NC-CC composition of the late veneer alone (and a NC composition of the Moon-forming impactor) is not sufficient to reproduce the BSE's Mo isotopic composition. Instead, a mixed NC-CC composition must be invoked for both the Moon-forming impactor and the late veneer. In this case, the late veneer and the Moon-forming impactor represent mixtures of genetically distinct materials, which combined produced the BSE's Ru and Mo isotopic compositions. Unlike in the previous model, no direct constraints on the specific Mo isotope anomalies of these two components can be deduced. A strength of this model is that a heterogeneous late veneer consisting of NC and CC material may resolve the discrepancy between evidence from refractory highly siderophile elements (i.e., Ru, Os), which are indicative of a NC-like late veneer (Fischer-Gödde and Kleine, 2017; Meisel et al., 1996), and evidence from the siderophile volatile elements Se, Te, and S, which suggest a carbonaceous chondrite-like late veneer (Wang and Becker, 2013; Varas-Reus et al., 2019). In this case, elements like Ru and Os would largely reflect the volatile-poor NC component, whereas elements such as Se, Te, and S would predominantly derive from the volatile-rich CC component in a NC-CC late veneer.

## 5. Conclusions

New Mo and Ru isotopic data for previously uninvestigated meteorite groups confirm the fundamental dichotomy between non-carbonaceous (NC) and carbonaceous (CC) meteorites observed in prior studies and extend the combined Mo-Ru isotopic data set available for NC meteorites significantly. With the exception of some meteorites from partially differentiated parent bodies, which display decoupled Mo and Ru isotope anomalies resulting from planetary processing of presolar carriers, the Mo and Ru isotope variations among NC meteorites are correlated as expected for the heterogeneous distribution of a common *s*-process carrier, consistent with prior work. Contrary to previous studies, we find that neither CC meteorites nor the BSE plot on the Mo-Ru correlation defined by NC meteorites. These observations demonstrate that the late-stage building blocks of Earth, including the Moon-forming impactor and the late veneer, consisted of a heterogeneous mixture of NC and CC materials. The two most likely scenarios involve either a NC late veneer combined with a CC Moon-forming impactor, or mixed NC-CC compositions for both components. The accretion of presumably volatile-rich CC material during the final stages of Earth's growth is consistent with chemical evolution models for core formation on Earth, which argue for heterogeneous accretion and the addition of more oxidized and volatile-rich material toward the end of Earth's formation (e.g., Wade and Wood, 2005; Rubie et al., 2011; Rubie et al., 2015). The results of the present study, therefore, resolve the inconsistencies between homogeneous accretion models inferred from prior interpretations of the Mo-Ru correlation, and the chemical evidence for heterogeneous accretion of Earth.

**Acknowledgments**

We gratefully acknowledge Addi Bischoff, Knut Metzler, the Institute of Meteoritics, University of New Mexico, and NASA for providing samples. US Antarctic meteorite samples are recovered by the Antarctic Search for Meteorites (ANSMET) program, which has been funded by NSF and NASA, and characterized and curated by the Department of Mineral Sciences of the Smithsonian Institution and Astromaterials Curation Office at NASA Johnson Space Center. We further thank Ulla Heitmann for technical support and sample preparation. This research received funding to Thorsten Kleine from the European Research Council under the European Community's Seventh Framework Program (FP7/2007-2013 Grant Agreement 616564 'ISO-CORE') and was supported by the Deutsche Forschungsgemeinschaft (SFB-TRR170, subproject B3). This is TRR 170 publication No. 85.